\documentclass[a4paper,11pt]{article}
\pdfoutput=1
\usepackage{jheppub}
\usepackage[T1]{fontenc}
\usepackage{amsfonts}
\usepackage{amsmath}
\usepackage{amssymb}
\usepackage{multirow}
\usepackage{graphicx}
\usepackage{mathtools}
\usepackage{comment}

\numberwithin{equation}{section}

\begin{document}

\title{Electric shocks: bounding Einstein-Maxwell theory with time delays on boosted RN backgrounds}

\author[a]{Sera Cremonini,}
\emailAdd{cremonini@lehigh.edu}
\affiliation[a]{Department of Physics, Lehigh University, Bethlehem, PA, 18015, USA}
\author[b,c]{Brian McPeak,}
\emailAdd{brian.mcpeak@df.unipi.it}
\affiliation[b]{Department of Physics, University of Pisa and INFN, Pisa, Italy}
\affiliation[c]{Department of Physics, McGill University, Montreal, Canada}

\author[a]{Yuezhang Tang}
\emailAdd{yut318@lehigh.edu}

\date{\today}

\abstract{The requirement that particles propagate causally on non-trivial backgrounds implies interesting constraints on higher-derivative operators. This work is part of a systematic study of the positivity bounds derivable from time delays on shockwave backgrounds. First, we discuss shockwaves in field theory, which are infinitely boosted Coulomb-like field configurations. We show how a positive time delay implies positivity of four-derivative operators in scalar field theory and electromagnetism, consistent with the results derived using dispersion relations, and we comment on how additional higher-derivative operators could be included. 

We then turn to gravitational shockwave backgrounds. We compute the infinite boost limit of Reissner-Nordstr{\"o}m black holes to derive charged shockwave backgrounds. We consider photons traveling on these backgrounds and interacting  through four-derivative corrections to Einstein-Maxwell theory.  The inclusion of gravity introduces a logarithmic term into the time delay that interferes with the straightforward bounds derivable in pure field theory, a fact consistent with CEMZ and with recent results from dispersion relations. We discuss two ways to extract a physically meaningful quantity from the logarithmic time delay -- by introducing an IR cutoff, or by considering the derivative of the time delay -- and  comment on the bounds implied in each case. Finally, we review a number of additional shockwave backgrounds which might be of use in future applications, including spinning shockwaves, those in higher dimensions or with a cosmological constant, and shockwaves from boosted extended objects.}

\maketitle

\section{Introduction}

Effective field theories (EFTs) describe physical systems in a maximally agnostic way: first, the low-energy degrees of freedom are identified, and then a theory is written down which includes every possible interaction involving those particular degrees of freedom, consistent with the symmetries. The couplings of these interactions, so called ``EFT coefficients,'' are determined by the high energy physics and can be computed explicitly when the latter is known. It is not possible to go the other way -- to derive the ultraviolet (UV) completion from its EFT description.

Despite the vast landscape of consistent UV theories, not every EFT can be UV completed \cite{Adams:2006sv}: there are certain choices of EFT coefficients 
which could never be obtained from 
a well-behaved high energy theory 
(which is unitary, causal, etc.). This can be clearly demonstrated using dispersion relations, where the assumption of analyticity of the amplitude plus bounds on the growth at large energies are used to write contour integrals in the $s$-plane relating the high and low energy parts \mbox{\cite{Pham:1985cr, Pennington:1994kc,  Ananthanarayan:1994hf, Comellas:1995hq, Dita:1998mh, Adams:2006sv}}. These have been studied extensively, and have been given new life recently due to the invention \cite{Arkani-Hamed:2020blm, Bellazzini:2020cot, Tolley:2020gtv, Caron-Huot:2020cmc, Sinha:2020win} of efficient algorithms for extracting bounds on the space of weakly coupled EFTs. These methods have been applied to a large number of scenarios~\cite{Manohar:2008tc, Mateu:2008gv, Nicolis:2009qm, Baumann:2015nta, Bellazzini:2015cra, Bellazzini:2016xrt, Cheung:2016yqr, Bonifacio:2016wcb,  Cheung:2016wjt, deRham:2017avq, Bellazzini:2017fep, deRham:2017zjm, deRham:2017imi, Hinterbichler:2017qyt, Bonifacio:2017nnt, Bellazzini:2017bkb, Bonifacio:2018vzv, deRham:2018qqo, Zhang:2018shp, Bellazzini:2018paj, Bellazzini:2019xts, Melville:2019wyy, deRham:2019ctd, Alberte:2019xfh, Alberte:2019zhd, Bi:2019phv, Remmen:2019cyz, Ye:2019oxx, Herrero-Valea:2019hde, Zhang:2020jyn, Trott:2020ebl,Zhang:2021eeo, Wang:2020jxr, Li:2021lpe, Du:2021byy, Davighi:2021osh, Chowdhury:2021ynh, Henriksson:2021ymi, Caron-Huot:2021enk, Caron-Huot:2022jli, Caron-Huot:2022ugt, Bern:2021ppb, Henriksson:2022oeu, Fernandez:2022kzi, Albert:2023jtd,Bellazzini:2023nqj, McPeak:2023wmq} and have proven extremely powerful, but their use has been plagued for decades by the same nagging question concerning the exact domain of analyticity of the $2 \to 2$ amplitudes. The bounds require maximal analyticity, which is generally thought to be a consequence of causality. For simpler cases this relationship is clear, but in quantum field theory the situation is more subtle. 
Understanding analyticity in full detail is an interesting and important research question. Still, it would be beneficial to have a parallel approach to bounding EFTs. 

The approach we shall take in this paper is to look at causality on classical backgrounds. In particular, we shall focus on photon propagation on backgrounds with boosted electric charges. It has been long known \cite{Aichelburg} that shockwave solutions arise from the infinite-boost limit of black hole backgrounds. Here we shall examine backgrounds of boosted charges with and without gravity, in order to probe the leading coefficients of Einstein-Maxwell and Maxwell theory, respectively. These can be summarized with the Lagrangian
\begin{align}
\label{introLag}
    \mathcal{L} \ = \ \sqrt{-g} \left( \frac{M_p^2}{2} R -\frac{1}{4} F_{\mu \nu} F^{\mu \nu} + \alpha_1 (F_{\mu \nu} F^{\mu \nu})^2 + \alpha_2 ( F_{\mu \nu} \tilde{F}^{\mu \nu})^2 + \alpha_3 W_{\mu \nu \rho \sigma} F^{\mu \nu} F^{\rho \sigma} \right),
\end{align}
with Maxwell theory arising in the $M_p \to \infty$ limit, in which case the $R$ and $W_{\mu \nu \rho \sigma} F^{\mu \nu} F^{\rho \sigma}$ terms decouple. Our results are the following:
\begin{enumerate}
    \item for Maxwell theory, causal propagation of photons on boosted charge backgrounds implies the following positivity bounds, 
    \begin{align}
        \alpha_1 \ > \ 0, \qquad \alpha_2 \ > \ 0 \, ,
    \end{align}
    in agreement with the dispersive bounds \cite{Bellazzini, Henriksson:2021ymi}.
    \item When gravity is included, 
   and the background is a four-dimensional boosted black hole with (rescaled) mass $m_0$ and charge $q_0$,  
    we find a time delay $\Delta v$ of the form
    \begin{align}
   \Delta v= -\frac{3 \pi q_0^2}{M_\text{P}^2 \rho} - 16 \frac{m_0}{M_\text{P}^2} \log\rho   
         +    \alpha_i  \frac{48 \pi q_0^2}{\rho^3} \pm  \,\alpha_3 
 \left(\frac{18 \pi q_0^2}{M_\text{P}^2 \rho^3} - \frac{64 m_0}{M_\text{P}^2 \rho^2} \right),
 \label{eq:intdelaywithmp}
\end{align}
\end{enumerate}
where $\rho$ is the impact parameter, and 
$\alpha_i$ can be $\alpha_1$ or $\alpha_2$, depending on the choice of polarization. Moreover, the $\alpha_3$ term can take either sign depending on whether the charge $q_0$ is taken to be electric or magnetic. Thus, the expression~(\ref{eq:intdelaywithmp}) incorporates 4 different time delays. It is difficult to interpret this equation in a straightforward way because of the presence of the $\log \rho$ term. In section~\ref{sec:Gravity} we discuss how a physically meaningful quantity can be extracted from this either by introducing a large $\rho$ / infrared cutoff, or by taking the derivative of the time delay with respect to $\rho$.

One of the primary technical difficulties in deriving this result is that the electric field vanishes in the infinite boost $\gamma \to \infty$ limit. More precisely, we are required to scale the mass and charge to zero as $m = m_0 \,\gamma^{-1}$ and $q^2 = q_0^2 \,\gamma^{-1}$ -- otherwise, the energy of the spacetime diverges at infinite boost. The gauge field $F = dA$, which is proportional to $q$, goes to zero in this limit. However we shall see that various quantities formed from contractions of $F$ do not, which will allow the higher derivative terms in equation~(\ref{introLag}) to source time delays.  

One of our motivations was to find bounds relevant for the black hole weak gravity conjecture \cite{Kats:2006xp}, which is the statement that the higher-derivative corrected black holes can be slightly super-extremal, having $q > m$ (see \cite{Jones:2019nev} for the generalization to magnetic charges). This is equivalent to an inequality on the four-derivative coefficients of Einstein-Maxwell theory -- in particular, $4 \alpha_1 \pm \alpha_3 / M_p^2 > 0$ for electric/magnetic black holes and $\alpha_2 > 0$ for dyonic (see also \cite{Cremonini:2020smy,Cremonini:2019wdk}).
It was pointed out \cite{Cheung:2014ega, Hamada:2018dde, Bellazzini:2019xts} that the combination of EFT coefficients relevant to the WGC appears in certain forward combinations of the $2 \to 2$ photon amplitude, leading to the hope that dispersion relations could be used to prove the conjecture. However, it is now understood that the pole due to graviton exchanges in those amplitudes invalidates the direct application of forward limit bounds. Methods were later developed \cite{Caron-Huot:2021rmr} for obtaining (weakened) positivity bounds in the presence of graviton exchanges; these methods were applied to photon amplitudes in \cite{Henriksson:2022oeu}, where the expected weakening of forward-limit positivity bounds was observed. The end result is a small allowable violation of the WGC which can be traced to a $\log$ of the impact parameter, exactly as we see above.

\paragraph{Causality bounds} Considering causality on classical backgrounds was one of the methods originally used by \cite{Adams:2006sv} to demonstrate that the leading higher derivative coefficient in scalar field theory is positive: one constructs a background where the higher-derivative corrections to the equations of motion affect the propagation speed of the scalar. The propagation will only remain causal if the scalar propagates more slowly than the speed of light, which requires a positive coefficient. This requirement was considered in \cite{Cheung:2014ega} in parallel with the dispersion relation bounds, where it was observed that forbidding superluminal propagation on certain backgrounds gives similar but slightly weaker bounds than the dispersive arguments. Recently there has been a revival of works using the requirement of causal propagation on non-trivial backgrounds to bound EFTs, including theories of scalars \cite{CarrilloGonzalez:2022fwg} and photons \cite{CarrilloGonzalez:2023cbf}. The current status appears to be that the causality bounds are strictly weaker than the dispersion relation bounds, though this may change when more general backgrounds are considered. Along these lines, EFTs in an expanding universe were considered in \cite{CarrilloGonzalez:2023rmc}. 

A conceptually similar idea was used in \cite{Camanho:2014apa}, referred to as ``CEMZ'', where propagation of photons and gravitons through shockwave backgrounds was analyzed in the presence of higher-derivative interactions. They found that Einstein gravity causes a universal time delay but the higher-derivative interactions will always cause a time advance for some choice of polarization. The conclusion was a problem with causality unless the gravitation time delay is larger than the higher-derivative time advance, implying a parametric bound on the relative size of the interactions. The aim here is mainly to generalize this to other charged backgrounds so that we can obtain bounds on the higher-derivative correction to Einstein-Maxwell theory. 

In theories of gravity, giving a precise physical interpretation of the time delay requires answering: delayed compared to what? It was shown in \cite{Gao:2000ga} that comparing the speed of light between a curved spacetime and a reference Minkowski background leads to gauge-dependent statements of causality. The authors were then able to prove a precise statement about causality in AdS by appealing to the boundary structure. In \cite{Camanho:2014apa}, an experimental procedure to define the time delay unambiguously by comparing with observers very far away in $d>4$ is given. This notably fails in $4d$ due to the $\log \rho$ term in the time delay. Since CEMZ, a number of papers \cite{Goon:2016une, Hinterbichler:2017qyt, deRham:2020zyh, AccettulliHuber:2020oou, Bellazzini:2021shn, deRham:2021bll, Chen:2021bvg,  Bittermann:2022hhy, Chen:2023rar} have used these and related ideas to investigate causality bounds on EFTs with gravity; see \cite{deRham:2022hpx} for a review.

\paragraph{Shockwaves and amplitudes}

The shockwave solutions to General Relativity were derived by Aichelburg and Sexl (AS) in 1971 by considering the infinite boost limit of a boosted black hole \cite{Aichelburg}. The resulting metric 
\begin{align}
    ds^2 \ = \ \eta_{\mu \nu} dx^\mu dx^\nu - 4 \, G \, m_0 \log\rho^2 \, \delta(u) \,  du^2 \, ,
\end{align}
describes a planar shockwave localized at $u = 0$, sourced by a particle moving in the $x$-direction at the speed of light -- see section~\ref{sec:Gravity} for a review. A key issue is that the total energy of the spacetime diverges in the infinite boost limit, unless the mass of the solution scales to zero as $M = m_0 \,  \gamma^{-1}$  
(here $\gamma$ is the Lorentz factor). Numerous later works extended this to other scenarios, which we partially review in section~\ref{sec:furthershockwaves}. These include the series of papers by Lousto and Sanchez \cite{Lousto:1988ua, Lousto:1989ha, Lousto:1990wn} who extended the AS solution to charged, spinning, and extended (non-pointlike) solutions. It is the charged solutions which shall be relevant for the present work.

Dray and 't Hooft \cite{Dray:1984ha} showed how the AS metric can be obtained by a coordinate change performed after $u = 0$, meaning that the AS solution can be thought of as two copies of Minkowski space ``stitched together'' along the null $u = 0$ plane. Later `t Hooft used this insight to derive the scattering amplitude of a particle traveling through the shockwave, which he then interpreted as a $2 \to 2$ scalar amplitude in the large $Gs$ regime:
\begin{align}
    A(s, t) = \frac{\Gamma(1 - i G s)}{4 \pi \Gamma(i G s)} \left( \frac{4}{-t} \right)^{1 - i G s} \, . 
\end{align}
The strikingly simple form of this amplitude has inspired a large amount of later work. 
A particularly important step forward was the computation of Kabat and Ortiz \cite{Kabat:1992tb} which showed how this amplitude arises from summing an infinite number of graviton-exchange ladder and crossed-ladder diagrams in the large $s / t$ limit. This remarkable result, in which an infinite number of loop diagrams resum into a \textit{classical} effect, gave a precise relation between the eikonal amplitudes and the shockwave backgrounds, and has spurred an enormous amount of effort on recovering classical physics from amplitude backgrounds, \textit{e.g.} \cite{Bjerrum-Bohr:2018xdl, Kosower:2018adc} for recent work and \cite{Buonanno:2022pgc, DiVecchia:2023frv} for recent reviews.

\paragraph{Conventions and important formulas}

In this paper, we shall use a mostly-plus metric convention, $\eta = \text{diag}(-1, 1, 1, ...)$. We shall also often make use of null coordinates $u = t - x_1$, $v = t + x_1$. In mostly-plus conventions, this leads to a flat space metric 
\begin{align}
    ds^2 \ = \ - du dv + dx_2^2 + dx_3^2 + ... \, .
\end{align}
We will typically consider four spacetime dimensions, and use $x = x_1$, $y = x_2$, and $z = x_3$.
Finally, an important formula that we will make use of throughout the paper is 
\begin{align}
\lim_{\epsilon \to 0} \frac{\epsilon^{n-\frac{1}{2}}}{(u^2 + 2 \epsilon \rho^2)^n}  \ = \  \frac{ \Gamma\left(n - \frac{1}{2}\right) }{\Gamma\left(n\right)} \frac{\pi^{1/2}}{(\sqrt2 \rho)^{2n - 1}} \delta(u)  \, .
    \label{eq:deltaformula}
\end{align}
For the special case of $n=\frac{1}{2}$, we use the following identity (see Section 3 for details),
\begin{align}
    \lim_{\epsilon \to 0}\frac{1}{\sqrt{u^2+2\epsilon\rho^2}}=-\log\rho^2 \delta(u) + \frac{1}{|u|} \,\,\,.
    \label{eq:deltaformulalog}
\end{align}

\paragraph{Outline of the paper}
In Section \ref{QFTsection} we discuss time delays in the presence of higher derivative operators, in theories without gravity.
Section \ref{sec:Gravity} discusses how to construct shockwaves by boosting black holes, and examines the time delays experienced by various probes propagating on these backgrounds,   in the presence of four-derivative corrections to Einstein-Maxwell theory. 
In Section \ref{sec:furthershockwaves} we compile a list of  additional shockwave geometries. Finally, we conclude in Section \ref{sec:Conclusions}.

\section{Shockwaves and time delays in field theory}
\label{QFTsection}

Let us start by considering time delays in theories without gravity.

\subsection{Scalar field}

The simplest possible example is a massless scalar field described by the Lagrangian
\begin{align}
    \mathcal{L} \ = \ -\frac{1}{2} (\partial \phi)^2  + \frac{g_2}{2} (\partial_\mu \phi)^4 + \frac{g_3}{3} (\partial_\mu \partial_\nu \phi)^2 (\partial_\rho \phi)^2 + 4 g_4 (\partial_\mu \partial_\nu \phi)^4 + ...
    \label{eq:scalar_Lagrangian}
\end{align}
For now we shall consider only the leading correction $g_2$ by assuming that it is much larger than the six- and higher-derivative corrections. The equations of motion are
\begin{align}
    \begin{split}
            0 \ = \ \partial^2 \phi - & 2 g_2 \,  \partial_\mu (\partial^\mu \phi \, (\partial \phi)^2 ) \, .
    \end{split}
    \label{eq:scalarEOMg2}
\end{align}
Our aim is to understand how the propagation of $\phi$ on some non-trivial backgrounds constrains the coefficients of the higher-derivative terms. We shall break $\phi$ into a background part $\phi_0$ and a probe $\pi$. The latter will be a small fluctuation so we will only consider its equations of motions to linear order.  The result of this is
\begin{align}
    \begin{split}
            \partial^2 \pi \ = \  2 g_2 \,  \partial_\mu (\partial^\mu \pi (\partial \phi_0)^2 + 2 \partial^\mu \phi_0 \partial_\nu \phi_0 \partial^\nu \pi ) \, . 
    \end{split}
    \label{eq:scalarEOMg2_linear}
\end{align}

To get a background that is  ``shockwave-like'', we start by taking $\phi_0$ to be a radially-pointing Coulomb-like field,
\begin{align}
    \phi_0 = \frac{k}{\sqrt{x^2 + y^2 + z^2}}\, ,
\end{align}
and then we boost along the $x$ direction,
\begin{align}
    t \to t' \ = \gamma (t - \beta x), \qquad x \to x' \ = \gamma (x - \beta t) \, .
\end{align}
After taking $\beta = 1 - \epsilon$ and expanding for small $\epsilon$, the equation of motion simplifies to
\begin{align}
 \partial_u \partial_v \pi \ = \ \frac{8 g_2 \, k^2 u \,\epsilon}{(u^2 + 2 \epsilon \rho^2)^3} \left(  4 \epsilon \partial_v \pi + (u - v \epsilon) \partial_v^2 \pi - 2 u \epsilon \partial_u \partial_v \phi \right) \, ,
 \label{eq:scalarEOM}
\end{align}
where we have assumed that the probe $\pi$ travels only in the $u = t - x$, $v = t+x$ plane, and defined $\rho = \sqrt{y^2 + z^2}$. This diverges in the $\epsilon \to 0$ limit. One way to handle this, which we shall have more to say about later, is to rescale $k^2 \to k_0^2 \gamma^{-1}$, where $k_0$ is fixed in the infinite boost limit. If we make this replacement, the result at small-$\epsilon$ becomes
\begin{align}
    \partial_u \partial_v \pi +  g_2 \frac{ \pi}{2} \frac{k_0^2}{\rho^3} \,\delta(u) \,\partial_v^2 \pi \ = \ 0 \, .
    \label{eq:g2EOM}
\end{align}
We see that this takes a directly analogous form to the equations of motion for the scalar traveling through a gravitational shockwave in \cite{Camanho:2014apa}. 
The solutions are oscillating functions with a time delay given by the coefficient of $\delta(u)$ in equation~(\ref{eq:g2EOM}), \textit{e.g.} of the form
\begin{align}
    \cos(v + \Delta v \, \theta(u)) \, ,
\end{align}
with the time delay given by
\begin{align}
    \Delta v  \ = \ g_2 \frac{\pi}{2} \frac{k_0^2}{\rho^3}  \, .
\end{align}
Note that the positivity of $\Delta v$ requires that the coefficient $g_2$ of the four derivative interaction in (\ref{eq:scalar_Lagrangian}) is positive. We see that the time delay falls off like $\rho^3$ -- this will be a general feature of four-derivative corrections.

It is also possible to compute the dependence on higher-derivatives, which contribute with higher inverse powers of $\rho$, such as
\begin{align}
    \Delta v  \ = \ g_2 \frac{\pi}{2} \frac{k_0^2}{\rho^3}  + g_3 \, 9 \pi \frac{k_0^2}{\rho^5} + ... \, .
\end{align}
We see that we are able to ignore the effect of $g_3$ and higher coefficients by considering very large distances $\rho$. To compare $g_2$ and $g_3$ would require that we look at the minimal distance, $\rho \sim 1 / M$. However this would lead to all coefficients $g_i$ contributing, requiring some other method to disentangle them, perhaps by considering more complicated potentials. It would be very interesting to try to understand how to do this systematically to reproduce the bounds of \cite{Caron-Huot:2020cmc}. The result would be a boosted version of \cite{CarrilloGonzalez:2022fwg}, which we believe could be very useful due to the large simplifications that occur in the infinite boost limit.

\subsection{Shockwaves in electromagnetism}

Photons provide another simple example. Maxwell theory is exactly linear so shockwave backgrounds have no effect on photon propagation. This means that the four-derivative corrections provide the leading interaction between the photons. Consider the Lagrangian
\begin{align}
    \mathcal{L} \ = \ -\frac{1}{4} F_{\mu \nu} F^{\mu \nu} + \alpha_1 (F_{\mu \nu} F^{\mu \nu})^2 + \alpha_2 ( F_{\mu \nu} \tilde{F}^{\mu \nu})^2  \, ,
\end{align}
with the corresponding equations of motion
\begin{align}
    \partial_\mu F^{\mu \nu}  = \partial_\mu \left( 8 \alpha_1 F^{\mu \nu}  F_{\alpha \beta} F^{\alpha \beta} + 8 \alpha_2 \tilde{F}^{\mu \nu}  F_{\alpha \beta} \tilde{F}^{\alpha \beta}  \right) \, .
\end{align}
One puzzle, pointed out in \cite{Lousto:1988ua}, is that the gauge field $F_{\mu \nu}$ vanishes if we use the scalings required to ensure finite-energy solutions in the ultrarelativistic limit $\epsilon \to 0$. However, we shall see that the time delay is still non-zero, as long as we compute the equations for a general boost and take the $\epsilon \to 0$ limit only at the end.

\subsubsection{Boosted solution}

We consider an 
electric monopole, whose potential is given by
\begin{align}
    A = -\frac{2q}{r} dt \, . 
\end{align}
This leads to a field strength
\begin{align}
    E(r) = F_{rt} = \frac{2q}{r^{2}}\, .
\end{align}
The magnetic field can be determined by dualizing this expression. Denoting the magnetic charge by $p$, the field strength for the entire configuration is then given by 
\begin{align}
    F_{\mu \nu} \ = \ \frac{2}{(x^2 + y^2 + z^2)^{3/2}}\begin{pmatrix}
    0 & -q x & - q y  & - q z \\
    q x & 0 & p z & -p y \\
    q y  & -p z  & 0 & p x  \\
     q z & p y  & -p y & 0 
    \end{pmatrix} \, .
\end{align}
Now, boosting with velocity $\beta$ in the $x$-direction and switching to $u$ and $v$ coordinates (but leaving the matrix in the $(t,x,y,z)$ basis) gives

\begin{align}
    F_{\mu \nu} \ = \ \frac{4 \epsilon}{(u^2 + 2 \epsilon \rho^2)^{3/2}}\begin{pmatrix}
    0 & q u & - q y - p z & p y -q z \\
    -q u & 0 & p z + q y &- p y+ q z\\
    q y +p z & -q y - p z & 0 & -p u  \\
     - p y+ q z & p y -q z  & p u & 0 
    \end{pmatrix} \, .
\end{align}
As a result, the square of the field strength is given by
\begin{align}
    F_{\mu \nu} F^{\mu \nu} =  \frac{32 (p^2 -q^2) \epsilon^2}{(u^2 + 2 \epsilon \rho^2)^2} \, .
\end{align}

\paragraph{Scaling the charges.}
Focusing for simplicity on the electric monopole case, note that if we assume that $q$ is constant as $\epsilon \to 0$, we obtain
\begin{align}
    F_{\mu \nu} \ = \ \frac{4 q}{\rho^2} \delta(u) \begin{pmatrix}
    0 & 0 & -y  & -z \\
    0 & 0 & y & z\\
    y  & -y  & 0 & 0  \\
     z & -z  & 0 & 0 
    \end{pmatrix} \, .
\end{align}
We see that the boosted solution is an electric field pointing radially out from the electric monopole, and a magnetic field curling around it, which is what we'd expect from a current. This solution gives us a finite electric and magnetic field in the infinite boost limit. However the energy of this field configuration diverges, as we can see by computing the stress tensor component $T_{00}$. One way to get a finite energy would be to scale the charge along with the boost. Finite energy requires that $p$ and $q$ scale as $\epsilon^{1/4}$ \cite{Lousto:1988ua}. Let us define 
\begin{equation}
    q^2 = \sqrt{2 \epsilon} \,
   q_0^2 \,, \qquad
    p^2 = \sqrt{2 \epsilon} \, p_0^2 \, ,
    \label{qpscalings}
\end{equation}
so that $q_0$ and $p_0$ are constant in the infinite boost limit. With this choice, we see that $F_{\mu \nu} \to 0$ and $F^2 \to 0$ as $\epsilon \to 0$. Nonetheless, there is still a finite time delay in this limit, as we will show below.

\subsubsection{Photons propagating on electric shockwaves}

Consider a probe photon $f_{\mu \nu}$ traveling on our shockwave background $F$. If we expand the equations of motion to first order in $f$, we find
\begin{align}
    \partial_\mu f^{\mu \nu} = 8 \alpha_1 \partial_\mu \left( f^{\mu \nu} (F^2) + 2 F^{\mu \nu} (F \cdot f) \right) + 8 \alpha_2 \partial_\mu \left( \tilde f^{\mu \nu} (F \cdot \tilde F) + 2 \tilde F^{\mu \nu} (f \cdot \tilde F) \right)\, .
\end{align}
The probe, described by the fluctuation $f$, is traveling in the $-x$ direction so that it crosses the shockwave, which is sourced by a charge traveling in the $+x$ direction. Before it hits the shockwave, the probe photon is propagating freely.
However, the higher-derivative terms act as a source altering the free motion of the photon, but the interaction  is localized at the $u = 0$ region of the shockwave.
In turn, this causes the probe to experience a time delay $\Delta v$, i.e. $v \to v + \Delta v$, when it crosses the shockwave. 

Moreover, the two polarizations of the photon will generically be \emph{mixed}
by the higher-derivative interaction. However, it is possible to choose the polarizations to conveniently ensure that only one of the higher-derivative terms contributes to the time delay of each mode, effectively ``diagonalizing'' the interaction. The choice which works is
\begin{align}
    A^{(1)}_\mu \ = \ \frac{\phi_1(u,v,y,z)}{y^2+ z^2} \left( - \frac{3}{2} \alpha_1 q_0 \, \theta(u) \, \Delta v   , \, \frac{3}{2}\alpha_1 q_0 \, \theta(u) \, \Delta v , \,  q_0 y + p_0 z , \, - p_0 y + q_0 z  \right) \\
    A^{(2)}_\mu \ = \ \frac{\phi_2(u,v,y,z)}{y^2+ z^2} \left( - \frac{3}{2} \alpha_2 p_0 \, \theta(u) \, \Delta v   , \, \frac{3}{2} \alpha_2 p_0 \, \theta(u) \, \Delta v , \,  p_0 y - q_0 z , \,  q_0 y + p_0 z  \right) .
    \label{eq:Aansatz}
\end{align}
Here we have used the rescaled charges introduced in  (\ref{qpscalings}).

Using the above ansatz, the equations of motion take a drastically simplified form: 
\begin{align}
    \partial_u \partial_v \phi_1 \ &= \ - \alpha_1 \,  \delta(u)  
    \Delta v \, \partial_v^2 \, \phi_1  \, , \\
    \partial_u \partial_v \phi_2 \ &= \ - \alpha_2 \, \delta(u) \,  \Delta v \,  \partial_v^2 \phi_2 \, , 
\end{align}
with the time delay given by
\begin{align}
    \Delta v \ = \ \frac{48 \pi (p_0^2 + q_0^2)}{ \rho^3} \, .
\end{align}
We see that $\alpha_1$ leads to a time delay in $\phi_1$ and $\alpha_2$ to a time delay in $\phi_2$. Thus, we conclude that 
\begin{align}
    \alpha_1 \, \geq \, 0 \, , \qquad \alpha_2 \, \geq \, 0 \, .
    \label{eq:EMbounds}
\end{align}
These are the same bounds on four-derivative operators which have been discussed from the point of view of dispersion relations: see \cite{Cheung:2014ega, Bellazzini:2019xts, Henriksson:2021ymi}, as well as \cite{Henriksson:2022oeu}, which observes a weakening of these bounds in the presence of gravity. In the next section, we shall turn to charged shockwaves on gravitational backgrounds, and see that the ``field theory bounds'' \eqref{eq:EMbounds} are indeed weakened when gravity is included.

\paragraph{Field strengths} Some properties of the solution are more clear when we examine the field strength that results from our choice of polarizations. For polarization 1, the electric fields are 
\begin{align}
    E_x \ &= \ - \alpha_1 \, \theta(u) \, \frac{3 q_0}{\rho^2} \Delta v \, \partial_v \phi_1 \, , \\
    E_y \ &= \ \frac{q_0 y + p_0 z}{\rho^2} \left( \partial_u \phi_1 + \partial_v \phi_1 \right)  +  \alpha_1 \, \theta(u)  \, \frac{15 q_0 y}{2 \rho^4} \Delta v \, \phi_1 \, ,\\
    E_y \ &= \  -\frac{p_0 y - q_0 z}{\rho^2} \left( \partial_u \phi_1 + \partial_v \phi_1 \right)   +  \alpha_1 \, \theta(u) \, \frac{15 q_0 z}{2 \rho^4} \Delta v \, \phi_1 \, .
\end{align}
When $p_0 = 0$ we see that this choice leads to the wave being polarized \textit{parallel} to $\rho = \sqrt{y^2 + z^2}$, while the polarization would be  \textit{perpendicular} to $\rho$ in the case where $q_0 = 0$.

We can also use this to compute the deflection angle due to the shockwave. For $u<0$, the electric field is pointing purely in the $y-z$ plane, with magnitude $\sqrt{p_0^2 + q_0^2} / \rho$. After the shock, the $x$-component of the electric field turns on, meaning that the wave has been deflected by an angle
\begin{align}
    \tan \theta_1 \ = \ \frac{-192 \pi  q_0 \sqrt{p_0^2 + q_0^2}}{\rho^4 } \, \alpha_1 \, .
\end{align}
Thus, we find that the deflection angle increases with the coupling $\alpha_1$ and with the amounts of charge $p_0$ and $q_0$, while it decreases with the distance from the shock.  A similar analysis applies to polarization 2 but with the role of $p_0$ swapped with $q_0$. In this case we find 
\begin{align}
    \tan \theta_2 \ = \ 
    \frac{-192 \pi  p_0 \sqrt{p_0^2 + q_0^2}}{\rho^4 } \, \alpha_2 \, .
\end{align}
In both cases, the electric field $E_x$ is non-vanishing after the shock but the magnetic field $B_x$ is still zero.

\section{Gravitational Shockwaves}
\label{sec:Gravity}

The goal of this section will be to boost Reissner-Nordstr\"om black holes to obtain charged shockwaves, and then to calculate  the time delays experienced by various probe particles propagating in the boosted background. We start by reviewing the derivation of the ultra-boosted Schwarzschild black hole for an example to illustrate how the calculation works.

\subsection{Review: shockwaves from boosted black holes}

First we will review the idea of \cite{Aichelburg}, where exact four-dimensional shockwave solutions were derived by considering the infinite boost limit of Schwarzschild black holes. We will work with a general number of dimensions $d$ first, and then consider $d = 4$, which is a special case due to the need to regulate logarithmic divergences which arise from the infinite boost limit. 

\paragraph{Boosted black holes}

Since the boost breaks spherical symmetry,  our starting point is the black hole in isotropic coordinates. We start with the usual Schwarzschild metric,
\begin{align}
    ds^2 \ = \ -\left( 1 - \frac{4 \kappa  m }{r^{d-3}} \right) dt^2 + \left( 1 - \frac{4 \kappa m }{r^{d-3}} \right)^{-1} dr^2 + r^2 d \Omega_{d-2}^2 \, , 
\end{align}
where
\begin{align}
    \kappa  \ = \ \frac{2 \pi \,  G \, \Gamma\left[ \frac{d-1}{2} \right]}{ (d - 2) \pi^{(d-1)/2}} \, .
\end{align}
After the change of coordinates
\begin{align}
    r \ \to \ r \left( 1 + \frac{\kappa m}{r^{d-3}} \right)^\frac{2}{d-3} \; ,
\end{align}
the metric takes the form
\begin{align}
    ds^2 \ = \ - \frac{\left( 1- \frac{\kappa m}{r^{d-3}} \right)^2 }{\left( 1+ \frac{\kappa m}{r^{d-3}} \right)^2} \, dt^2 + \left( 1 + \frac{\kappa m}{r^{d-3}} \right)^\frac{4}{d-3}\left( dr^2 + r^2 d\Omega_{d-2}^2\right)  \, ,
\end{align}
which is precisely in isotropic coordinates, since  
$\left( dr^2 + r^2 d\Omega_{d-2}^2\right) = \sum dx_i^2$.
Next, we boost along the $x$ direction,
\begin{align}
\label{section3boost}
    t \to t' \ = \gamma (t - \beta x), \qquad x \to x' \ = \gamma (x - \beta t) \, ,
\end{align}
and expand around the infinite boost limit, $\beta \to 1 - \epsilon$, with $\epsilon << 1$. Boosting a finite-mass particle to the speed of light will result in a diverging energy, so we simultaneously scale down the mass by
\begin{align}
    m \ \to \  \gamma^{-1} m_0  \ = \  \sqrt{2 \epsilon} \, m_0 + \mathcal{O}(\epsilon^{3/2}) \, .
\end{align}
Introducing $u=t-x$,
the resulting metric can be written in terms of flat space plus the leading correction at small-$\epsilon$:
\begin{align}
    ds^2 \ = \ \eta_{\mu \nu} dx^\mu dx^\nu + \kappa \, m_0 \,  \frac{ 2^{d/2} (d - 2) \, \epsilon^{(d-4)/2}}{(d-3)(u^2 + 2 \epsilon \rho^2)^{(d-3)/2}} \, du^2 \, .
\end{align}

\paragraph{Shockwaves in $d>4$: }

For $d>4$, we can take the small-$\epsilon$ limit of the metric above using formula~\eqref{eq:deltaformula}, which gives 
\begin{align}
    ds^2 \ = \ \eta_{\mu \nu} dx^\mu dx^\nu + \frac{4  \, G \,  m_0 \,\Gamma\left[\frac{d-4}{2} \right]}{\pi^{(d-4)/2} \rho^{d-4} } \delta(u) \, du^2 \, ,
    \label{eq:ASmetdg4}
\end{align}
in agreement with \cite{Camanho:2014apa}. 
As we shall show below, from this metric we can immediately read off the time delay of a scalar moving through the shockwave. The result is  
\begin{align}
    \Delta v \ = \ \frac{4 \, m_0 \, G \, \Gamma\left[\frac{d-4}{2} \right]}{\pi^{(d-4)/2} \rho^{d-4} } \, .
\end{align}

\paragraph{Special case of $d=4$}

Let us now consider the special case corresponding to $d=4$. Note that  the expression for the metric~\eqref{eq:ASmetdg4} 
is singular in $d=4$ because the $\Gamma$-function diverges. To see where this is coming from exactly, consider the finite-$\epsilon$ metric,
\begin{align}
    ds^2 \ = \ \eta_{\mu \nu} dx^\mu dx^\nu + \frac{4 \, m_0 \, G}{\sqrt{u^2 + 2 \epsilon \rho^2}} \, du^2 \, .
\end{align}
We can see that the  $du^2$ component does conform to our formula~\eqref{eq:deltaformula}, but the RHS is divergent because the Gamma-function in the numerator would be evaluated at zero\footnote{A clear way to see that we do not get a simple $\delta(u)$ factor is to integrate the metric component before taking the $\epsilon \to 0$ limit. This gives
\begin{align}
    \int_{-R}^R du \frac{1}{\sqrt{u^2 + 2 \epsilon \rho^2}} \ = \ -  \log \frac{\epsilon \rho^2}{2 R^2}
\end{align}
which is valid if we take $\epsilon \to 0$ \textit{first}. In addition to the small-$\epsilon$ limit arising from the large-boost limit, we also see that this diverges if $R$ or $\rho$ go to infinity.}. This is handled in \cite{Aichelburg} using a coordinate transformation\footnote{This redefinition is special to $d = 4$ and is responsible for the interesting relationship between shockwave backgrounds and gravitational memory, see \cite{He:2023qha}.}
\begin{align}
    x - \beta t \ &= \ x - \beta t \, , \\
    x + \beta t \ &= \ x + \beta t - 4 m_0 \log \left( \sqrt{(x - \beta t)^2 + (1 - \beta^2) } - (x - t) \right) \, ,
\end{align}
which takes the metric to the form 
\begin{align}
    ds^2 \ = \ \eta_{\mu \nu} dx^\mu dx^\nu + 4 \, G \, m_0 \left( \frac{1}{\sqrt{u^2 + 2 \epsilon \rho^2}} - \frac{1}{\sqrt{u^2 + 2 \epsilon}}\right)\, du^2 \, .
\end{align}
This removes the divergence in the $\epsilon \to 0$ limit, and the metric becomes 
\begin{align}
    ds^2 \ = \ \eta_{\mu \nu} dx^\mu dx^\nu - 4 \, G \, m_0 \log\rho^2 \, \delta(u) \,  du^2 \, .
\end{align}
Performing the inverse field-redefinition, as is done in \cite{Aichelburg}, to go back to the original coordinates yields
\begin{align}
    ds^2 \ = \ \eta_{\mu \nu} dx^\mu dx^\nu - 4 \, G \, m_0 \left( \log\rho^2 \, \delta(u) - \frac{1}{|u|}  \right) \,  du^2 \, .
\end{align}

A common claim in the literature (\textit{e.g.} \cite{tHooft:1987vrq}) is that this is equivalent to gluing two copies of Minkowski together, with
\begin{align}
    t_+ \ &= \ t_- + 2 \, G \, m_0 \log (\rho^2/C)  \\
    x_+ \ &= \ x_- + 2 \, G \, m_0 \log (\rho^2/C)  \\
    y_+ \ &= \ y_- \\
    z_+ \ &= \ z_- \, ,
\end{align}
where $C$ is an irrelevant constant. This redefinition has measurable effects, \textit{e.g.} two particles located at the same value of $x$ but different values of $\rho$ before the shockwave will have different values of $x$ after the shockwave (see \cite{Dray:1984ha} for a nice discussion and illustration).

\subsubsection{Time delays on black hole backgrounds}

Before moving on to the more complicated backgrounds of interest to us, let us review how to extract the time delay from the equation of motion of a scalar field propagating in a generic shockwave background, following \cite{Camanho:2014apa}. For cases where the scalar field
is minimally coupled to gravity, the geometry entirely determines the time delay, and the latter can be easily read off from the metric. This is because, for the general shockwave background 
\begin{align}
\label{flatsw}
    ds^2=-dvdu+h(u,x_i)du^2+\sum_{i=1}^{D-2}(dx_i)^2\,,
\end{align}
the wave equation for a scalar probe traveling through it is given by 
\begin{align}
\label{scalareom}    \nabla^2\phi=\partial_u\partial_v\phi+h\partial_v^2\phi-\frac{1}{4}\partial_i^2\phi=0\,\,.
\end{align}
We assume that the probe is only traveling in the $x$-direction, meaning that it does not depend on $y$ and $z$. Then the equation of motion becomes
\begin{align}
\partial_v\bigl(\partial_u\phi+h\partial_v\phi\bigr)=0 \, ,
\label{eq:scalarEOM2}
\end{align}
which mirrors the form of the EOM in equation~(\ref{eq:scalarEOM}). Now, if the shockwave profile can be modeled by $h = X \delta(u)$, then the solutions will be oscillating functions away from $u = 0$, and discontinuous at $u = 0$. To see this, we consider a solution $\phi^+$ to the free equation valid for $u>0$ and a solution $\phi^-$ valid for $u<0$, patched together at $u = 0$ according to the equation of motion. Specifically, consider the ansatz
\begin{align}
\label{28}
    \phi(u=0^+,v,x^i)=e^{-\int_{0^-}^{0^+}du\,h\, \partial_v}\phi(u=0^-,v,x^i)\,\,\,.
\end{align}
Indeed, this is a solution to (\ref{eq:scalarEOM2}), since acting on both sides with $\partial_u$ yields
\begin{align}
   \partial_u\phi(u=0^+,v,x^i)
    &=h \, \partial_v\phi(u=0^+,v,x^i)\,.
\end{align}
The ansatz above can be rewritten by shifting $v$ in the original solution, 
\begin{align}
   e^{-\int_{0^-}^{0^+}du\,h\, \partial_v}\phi(u=0^-,v,x^i) = e^{-X \,\partial_v}\phi(u=0^-,v,x^i) = \phi(u=0^-,v - X,x^i)\, ,
\end{align}
which shows that the time delay is indeed given by $X$. Thus, we have seen that the latter can be easily read off from either 
the metric (\ref{flatsw}) or from (\ref{scalareom}). 

\subsection{Boosted Reissner-Nordstr{\"o}m  black holes}

In four dimensions, the
Reissner-Nordstr{\"o}m  black hole can have both electric and magnetic charges, $q$ and $p$ respectively. In isotropic coordinates the metric is given by
\begin{align}
    d s^2=-g_{00}(\Bar{r}) \, dt^2+g_{11}(\Bar{r})\,(dx^2+dy^2+dz^2) \,,
\end{align}
where
\begin{align}
    g_{00}(\bar r) &=1-\frac{2m}{r}+\frac{q^2 + p^2}{r^2}\,, \qquad 
    g_{11}(\bar r) =\left( \frac{r}{\bar r}\right)^2 \,,
\end{align}
and $r$ is related to $\Bar{r}=\sqrt{x^2+y^2+z^2}$ through 
\begin{align}
    r=\Bar{r}\left(1+\frac{m}{\Bar{r}}+\frac{m^2-q^2 - p^2}{4\Bar{r}^2}\right).
\end{align}
The background gauge field which supports the geometry is given by 
\begin{align}
    A \ = \ \frac{2q}{r}dt+2p\cos{\theta}d\phi \ = \ \frac{2q}{r}dt +  \frac{2 p x}{\bar r} \frac{z dy - y dz}{y^2+z^2} \,.
\end{align}

Under the boost
\begin{equation}
    t \rightarrow \gamma (t - \beta x)\, , \quad 
    x \rightarrow \gamma (x - \beta t)\, , 
\end{equation}
the metric becomes
\begin{align}
   g_{\mu\nu}=\begin{pmatrix}
    \gamma^2(-g_{00}+\beta^2g_{11}) & \beta\gamma^2(g_{00}-g_{11}) & 0 & 0 \\
    \beta\gamma^2(g_{00}-g_{11}) & \gamma^2(-\beta^2g_{00}+g_{11}) & 0 & 0 \\
    0  & 0  & g_{11} & 0  \\
     0 & 0  & 0 & g_{11} 
    \end{pmatrix} \, ,
\end{align}
and the gauge field becomes
\begin{align}
    A = \frac{8 q \bar r \gamma}{m^2 - p^2 - q^2 + 4 m \bar r + 4 \bar{r}^2} (dt  - \beta dx) 
        -\frac{2 p (x- \beta t) \gamma}{\bar r (y^2 + z^2)}  (z dy - y dz) \, ,
\end{align}
with $\Bar{r}=\sqrt{\gamma^2 (x-\beta t)^2 + y^2 +z^2}$.

In the ultra-relativistic limit $\beta= 1-\epsilon$ with $\epsilon <<1$, 
after rescaling the mass and electric charge parameters of the black hole,
\begin{equation}
\label{RNrescalings}
    m \rightarrow m_0 \, \gamma^{-1} \, ,  \quad q^2 \rightarrow q_0^2 \,\gamma^{-1} \, , \quad
    p^2 \rightarrow p_0^2 \, \gamma^{-1} \, ,
\end{equation}
we find
\begin{equation}
    ds^2 = \eta_{\mu\nu}dx^\mu dx^\nu + \frac{1}{\sqrt{2 \epsilon}}   
    \left( \frac{4 m_0 }{\Bar{r}} - 
    \frac{3}{2}
    \frac{
    q_0^2+p_0^2 }{\Bar{r}^2}\right) du^2\, . 
\end{equation}
Using the identities given in the introduction for the infinite boost limit, we have
%
%
\begin{align}
\label{finalRN}
    ds^2 \ = \ \eta_{\mu \nu} dx^\mu dx^\nu - 
    \Biggl(8\,m_0\,\ln{\rho+\frac{3}{2}\pi\,\frac{q_0^2+p_0^2}{\rho}}\Biggr)\delta(u)
    \, du^2 \, .
\end{align}
See section~\ref{sec:furthershockwaves} for the higher-dimensional version of these results.

\subsubsection{Time delay of a charged scalar field}

For minimally coupled particles the time delay is determined entirely by the geometry. Let us see how this works for a charged scalar field. This example will allow us to make an important point about the difference between gauge fields and gravitational fields. 

We consider a complex scalar field $\Phi$ with charge $e$ and mass $M$ coupled to gravity and a $U(1)$ gauge field $A$, whose  
equation of motion is given by
\begin{align}
    -(\nabla_\mu-ieA_\mu)(\nabla^\mu+ieA^\mu)\Phi+M^2\,\Phi=0\, .
\end{align}
In the infinite boost limit $\beta = 1 - \epsilon$, with $\epsilon \ll 1$, it becomes 
\begin{align}
 \partial_u \partial_v \Phi -\left(\frac{3 \pi}{2}\frac{p_0^2 + q_0^2}{\rho} + 8 m_0 \log \rho \right) \delta(u) \partial_v^2 \Phi  + 4 \frac{m_0}{|u|} \partial_v^2 \Phi - \frac{1}{4}(\partial_y^2 + \partial_z^2 + M^2) \Phi = 0 \, .
\label{eq:chargedscalarEOM}
\end{align}
Note that the $m_0 / |u|$ term is the same correction that appears in the metric after switching back to the original coordinate system. Since only the first two terms
in (\ref{eq:chargedscalarEOM}) matter near $u = 0$, we can read off the time delay by comparing to (\ref{eq:scalarEOM2}), and we find
\begin{align}
   \Delta v= -\biggl(\frac{3\pi}{2}\frac{p_0^2+q_0^2}{\rho}+8m_0    \,\log\rho\biggr)\,.
\end{align}
So the time delay is exactly what we would have predicted from the geometry. This might be a surprising conclusion: the charge $e$ of the scalar field does not appear in the expression. The black hole's electric and magnetic fields do appear because they warp the geometry, but they do not interact with the scalar's charge in the infinite boost limit. This is quite reminiscent of the result from dispersion relations, where the electric charge does not contribute to doubly subtracted dispersion relations. It can, however, appear in singly subtracted dispersion relations -- see \cite{McPeak:2023wmq} for an analysis of electric charges and singly subtracted dispersion relations in the presence of gravity. Possibly a different choice of the charge scalings would have allowed $e$ to appear in the time delay -- it would be very interesting to understand better why non-dispersive coefficients also do not contribute to the time delay.

\subsection{Time delays in theories with higher derivatives}

Next, we are interested in examining time delays in Einstein-Maxwell theories with four-derivative corrections of the form,
\begin{align}
    \label{eq:EMlagrangian} 
    S \ = \ \int  d^D x \, \sqrt{-g}\,\left(R-\frac{1}{4} F_{\mu \nu} F^{\mu \nu} + \alpha_1 (F_{\mu \nu} F^{\mu \nu})^2 + \alpha_2 ( F_{\mu \nu} \tilde{F}^{\mu \nu})^2 
    + \alpha_3 W^{\mu \nu \alpha \beta}F_{\mu \nu}F_{\alpha \beta}\right) \, ,
\end{align}
where we set $M_P^2=2$ and it is understood that the parameters $\alpha_i$ are perturbatively small. 
We will focus on four spacetime dimensions.
We want to follow the propagation of a probe photon on a shockwave background corresponding to a boosted charged black hole. 
To this end, we add a small 
fluctuation to the gauge field,
$F_{\mu\nu}+ f_{\mu\nu}$, and the resulting equation of motion is given by 
\begin{align}
\label{photonEOM}
    \nabla_\mu f^{\mu \nu} = & 8 \alpha_1 \,\nabla_\mu \left( f^{\mu \nu} F^2 + 2 F^{\mu \nu} \, F \cdot f \right) + 8 \alpha_2 \,\nabla_\mu \left( \tilde f^{\mu \nu} \,F \cdot \tilde F + 2 \tilde F^{\mu \nu} \,f \cdot \tilde F \right) \nonumber \\ & + 4\alpha_3 \, \nabla_\mu \left(W^{ \mu \nu \alpha \beta}f_{\alpha \beta} \right)\, .
\end{align}
We are going to choose the photon polarizations to be given as in (\ref{eq:Aansatz}).

\paragraph{Equations of motion} 

For simplicity we examine first the case in which the photon is propagating on a purely electrically charged shockwave solution, i.e. the magnetic charge $p_0$ is turned off.
The time delay can be extracted from either one of the bottom two components of (\ref{photonEOM}). 
For polarization one, 
we find 
\begin{align}
        &   
           \partial_u \partial_v \phi + \frac{ 4 m_0}{|u|} \partial^2_v \phi + \left[ -\frac{3 \pi q_0^2}{2\rho} - 8 m_0 \log\rho   
         +    \alpha_1  \frac{48 \pi q_0^2}{\rho^3} +  \,\alpha_3 
 \left(\frac{9 \pi q_0^2}{ \rho^3} - \frac{32 m_0}{\rho^2} \right)\right] \delta(u) \, \partial^2_v \phi  = 0,
\end{align}
while for polarization two
we have
\begin{align}
        &   
           \partial_u \partial_v \phi + \frac{ 4 m_0}{|u|} \partial^2_v \phi + \left[ -\frac{3 \pi q_0^2}{2\rho} - 8 m_0 \log\rho   
         +    \alpha_2  \frac{48 \pi q_0^2}{\rho^3} -  \,\alpha_3 
 \left(\frac{9 \pi q_0^2}{ \rho^3} - \frac{32 m_0}{\rho^2} \right)\right] \delta(u) \, \partial^2_v \phi  = 0 \, .
\end{align}

\paragraph{Time delays for electric black holes}

Once again, we can read off the time delays by comparing the structure of these equations to that of (\ref{eq:scalarEOM2}). For the polarization one mode, the time delay is
\begin{align}
   \Delta v= -\frac{3 \pi q_0^2}{2\rho} - 8 m_0 \log\rho   
         +    \alpha_1  \frac{48 \pi q_0^2}{\rho^3} +  \,\alpha_3 
 \left(\frac{9 \pi q_0^2}{ \rho^3} - \frac{32 m_0}{\rho^2} \right),
\end{align}
while for the polarization two mode we have
\begin{align}
   \Delta v= -\frac{3 \pi q_0^2}{2\rho} - 8 m_0 \log\rho   
         +    \alpha_2  \frac{48 \pi q_0^2}{\rho^3} -  \,\alpha_3 
 \left(\frac{9 \pi q_0^2}{ \rho^3} - \frac{32 m_0}{\rho^2} \right) .
\end{align}
As expected, the time delays  receive two distinct contributions, one purely from the geometry, which matches precisely what would have been extracted from the metric (\ref{finalRN}), and one from the higher derivative terms. 
Moreover, the contribution from the geometry weakens the bounds, as it comes in with the opposite sign.
Finally, note that the polarization one mode is only sensitive to the $\alpha_1$ correction, while the
polarization two mode is only sensitive to the $\alpha_2$ term. Indeed, the two polarizations were constructed so that this would be the case. On the other hand, they are both sensitive to $\alpha_3$, but \emph{with different signs}. 

\paragraph{Magnetic black holes}

We can also consider a shockwave created from boosting a purely magnetic black hole, i.e. turning off $q_0$ but keeping $p_0$. For the polarization one mode, the time delay is now given by
\begin{align}
   \Delta v= -\frac{3 \pi p_0^2}{2\rho} - 8 m_0 \log\rho   
         +    \alpha_1  \frac{48 \pi p_0^2}{\rho^3} - \,\alpha_3 
 \left(\frac{9 \pi p_0^2}{ \rho^3} - \frac{32 m_0}{\rho^2} \right),
\end{align}
while for the polarization two mode we have
\begin{align}
   \Delta v= -\frac{3 \pi p_0^2}{2\rho} - 8 m_0 \log\rho   
         +    \alpha_2  \frac{48 \pi p_0^2}{\rho^3} +  \,\alpha_3 
 \left(\frac{9 \pi p_0^2}{ \rho^3} - \frac{32 m_0}{\rho^2} \right) .
\end{align}
Thus, we see that the purely magnetic case can be obtained from the electric one, by sending $q_0$ to $p_0$ and at the same time swapping the sign of the $\alpha_3$ contribution in each of the two polarizations.
This already hints at the fact that the Weyl contribution to the higher derivative terms causes the polarizations to rotate among each other, as we are about to see more clearly below.

\paragraph{Dyonic black holes} Finally, we keep both electric and magnetic charges, i.e. consider shockwaves formed by boosting  dyonic black holes. 
For the polarization one mode, component 3 of (\ref{photonEOM}) yields
\begin{align}
        &(q_0 y + p_0 z) \, \partial_v \left[ \partial_u \phi +\frac{4 m_0}{|u|} \partial_v \phi -  \left(  8 m_0 \log \rho + (p_0^2 + q_0^2) \left(\frac{3 \pi }{2 \rho} - \frac{48 \pi \alpha_1}{\rho^3} \right)\right) \delta(u) \, \partial_v \phi \right] \nonumber \\
          & \qquad \qquad + (q_0 y - p_0 z) \, \alpha_3 \left(\frac{18 \pi (p_0^2 + q_0^2)}{\rho^3} - \frac{32 m_0}{\rho^2} \right) \delta(u) \, \partial_v^2 \phi = 0 \, ,
\end{align}
while component 4 gives
\begin{align}
        &(q_0 z - p_0 y) \, \partial_v \left[ \partial_u \phi +\frac{4 m_0}{|u|} \partial_v \phi -  \left(  8 m_0 \log \rho + (p_0^2 + q_0^2) \left(\frac{3 \pi }{2 \rho} - \frac{48 \pi \alpha_1}{\rho^3} \right)\right) \delta(u) \, \partial_v \phi \right] \nonumber \\
          & \qquad \qquad + ( q_0 z + p_0 y) \alpha_3 \, \left(\frac{9 \pi (p_0^2 + q_0^2)}{\rho^3} - \frac{32 m_0}{\rho^2} \right) \delta(u) \, \partial_v^2 \phi = 0 \, .
\end{align}
By inspecting the detailed dependence on the charges and the $y,z$ coordinates in the expressions above, it is clear that we cannot solve both equations. We have not included the equations for the second polarization, but the same issue appears. This issue would disappear if we were to set $\alpha_3=0$. 

The problem is that the term containing the Weyl tensor rotates the polarizations among each other in a quite non-trivial way. As a result it splits the incoming polarization into a combination of the two polarizations, with each mode experiencing a different time delay. Thus, it is clear that we cannot solve the equations of motion with our ansatz -- we have tried but have been unable to find a different ansatz that maintains the incoming polarizations. Whether this is just an inconvenience that requires a smarter choice of ansatz, or a fundamental issue, is not clear to us.

\subsection{Time delays and causality bounds}

Let us now comment on the meaning of these delays. First, let us restore the factors of $M_\text{P}$, which can be done using dimensional analysis and recalling that we set $M_\text{P}^2 = 2$, see e.g.  (\ref{eq:EMlagrangian}). The expression incorporating all possible delays is of the form
\begin{align}
   \Delta v= -\frac{3 \pi q_0^2}{M_\text{P}^2 \rho} - 16 \frac{m_0}{M_\text{P}^2} \log\rho   
         +    \alpha_i  \frac{48 \pi q_0^2}{\rho^3} \pm  \,\alpha_3 
 \left(\frac{18 \pi q_0^2}{M_\text{P}^2 \rho^3} - \frac{64 m_0}{M_\text{P}^2 \rho^2} \right),
 \label{eq:delaywithmp}
\end{align}
where $\alpha_i$ can be $\alpha_1$ or $\alpha_2$, and $q_0$ might be either the electric or magnetic charge. We immediately see that every term except for the $\alpha_i$ term decouples when $M_\text{P} \to \infty$ limit. This is consistent with the dispersion relation bounds \cite{Henriksson:2021ymi, Henriksson:2022oeu}.

There are two issues that make deriving precise bounds from the above expression a bit tricky. The first is to understand for which values of $\rho$ the above expression should hold. A naive application of the bounds for any $\rho$ will imply $\alpha_1 > 0$ and $\alpha_2 > 0$ because these terms dominate at very small $\rho$ (and because we can take appropriate linear combinations of the time delays to cancel the $\alpha_3$ dependence). Clearly this is not valid though, since at very short distances, the time delay will not be computable in the EFT. Thus, we will follow \cite{Camanho:2014apa} and use
\begin{align}
    \rho \gtrsim \frac{1}{M} \, ,
\end{align}
where $M$ is the cutoff scale of the EFT.

The second issue, which is particular to 4d, is the $\log \rho$ term. This is not well-defined without a reference scale. Below we shall consider two ways around this issue, both of which inspired by the principle that there are no ``short-cuts through the bulk''\footnote{This is also the spirit of the second Gao-Wald theorem, which states that for hyperbolic spacetimes, the geodesics at the conformal boundary are the fastest. See also \cite{Bittermann:2022hhy}, where a similar ``no shortcuts'' principle was applied to give sensible causality bounds in de Sitter space.}. In $d>4$, it is easy to see what this means -- every term in the time delay vanishes at $\rho = \infty$, so a positive time delay is equivalent to having a larger time delay at finite $\rho$ than at infinite $\rho$ \cite{Camanho:2014apa}. For $d = 4$ the time delay diverges at infinite $\rho$. One possibility is to choose a cutoff at a large value $\rho = \rho_0$. This would essentially mean imposing an IR regulator. Another possibility is to enforce the ``no shortcuts'' principle locally in the bulk, meaning that we require that the derivative of the time delay with respect to $\rho$ is always negative. Neither possibility is perfectly satisfactory: the cutoff idea introduces the arbitrary choice of $r_0$, and while the derivative is well defined, it is not clear why it would need to be positive\footnote{Once again we appeal to similarity with the Gao-Wald requirement that the boundary contains the fastest null geodesic. However it may be that this derivative can be proven to be positive through its relation to the deflection angle or some other quantity. We leave exploring this to future work.}. Nonetheless we shall consider both possibilities in turn and see what they would imply.

\paragraph{Bounds compared to infinity}

One way to obtain a measurable quantity from the time delay in (\ref{eq:delaywithmp}) is to compare two values. In $d>4$ this is easily done by comparing with $\rho = \infty$, where the delay vanishes. In $d = 4$ the time delay associated with the log term diverges at infinity, we shall instead choose a cutoff $r_0 \sim 1 / m_{\text{IR}}$ to compare to, and consider $r_0$ very large so that inverse powers of $r_0$ are essentially zero. If we require that the time delay for $\rho < r_0$ is always greater than the delay at $r_0$, this gives us 
\begin{align}
   \alpha_i    > - 16 \rho^3\frac{m_0}{3 \pi q_0^2 M_\text{P}^2} \log (r_0 / \rho  )  + \rho^2 \frac{1}{16 M_\text{P}^2 } \, , 
\end{align}
where we have canceled the $\alpha_3$ dependence by adding the $+ \alpha_3$ with the $-\alpha_3$ equations.

This equation can be directly compared to the results from dispersion relations \cite{Henriksson:2022oeu}, where it was observed that 
\begin{align}
   \alpha_1 \, > \,    - \frac{ c_1}{ M^2 M_\text{P}^2} \log ( M /  m_{\text{IR}})  + \frac{c_0}{M^2 M_\text{P}^2} \, .
\end{align}
The values of $c_1 = 24.26$ and $c_0 = 33.33$ were determined through the numerical optimization procedure introduced in \cite{ Caron-Huot:2021rmr, Caron-Huot:2022ugt}. Interestingly, if we have
\begin{align}
    \frac{m_0}{q_0^2} \ll M \, .
\end{align}
then we find that the causality bounds will be parametrically stronger than the dispersion relation bounds. Recall that $m$ and $q^2$ were both scaled by a factor of $\gamma^{-1}$, so $\frac{m_0}{q_0^2} = \frac{m}{q^2}$. We do not know of any type of bounds relating mass to charge squared; it would be interesting to understand if this is possible. 

\paragraph{Bound on the derivative} Another possible route around the issue that the time delay in 4d is not measurable is to consider its derivative. This was pointed out by Dray and `t Hooft \cite{Dray:1984ha}, who observed that after the shockwave passes, the observer cannot figure out how much the shockwave set back her clock. In $d > 4$, it is possible to compare with a ``clock at infinity,'' where the time delay has no effect, but this is not possible in 4d due to the log. However it \textit{is} possible to determine the rate of change of the time delay, by simply comparing clocks with a nearby observer. Hence the derivative is completely well-defined. This is given by
\begin{align}
    \frac{\partial \Delta v}{\partial \rho} \ = \ \frac{3 \pi q_0^2}{M_\text{P}^2 \rho^2} - 16 \frac{m_0}{M_\text{P}^2 \rho}   
         -    \alpha_i  \frac{144 \pi q_0^2}{\rho^4} \pm  \,\alpha_3 
 \left(\frac{56 \pi q_0^2}{M_\text{P}^2 \rho^4} - \frac{128 m_0}{M_\text{P}^2 \rho^3} \right).
 \label{eq:derdelay}
\end{align}

A few consequences are easy to see. In the absence of higher-derivative terms the negativity of this derivative implies 
\begin{align}
    \frac{m_0}{q_0^2} > \frac{3 \pi}{16 \rho} \gtrsim  M \, ,
\end{align}
where we have used the fact that the strongest possible bound comes from the smallest possible value of $\rho$, which is $\rho\sim 1/M$.
This indicates that the momentum $m_0$ cannot be arbitrarily small, as this would presumably take us out of the regime of eikonal scattering. We can also immediately see that in the limit $M_\text{P} \to \infty$, where gravity decouples, we find $\alpha_1 > 0$ and $\alpha_2 > 0$, as we should.

Next let us restore the higher-derivative terms but remove $\alpha_3$, again by adding the $+ \alpha_3$ equation to the $-\alpha_3$ equation. Then we have
\begin{align}
     \alpha_i  \ > \ \frac{1}{144 \pi  M_\text{P}^2}  \left(3 \pi  \rho^2 - 16 \frac{m_0}{q_0^2} \rho^3 \right) \, .
\end{align}
If the charge $q_0$ goes to zero, then the bound gets infinitely weak, consistent with the expectation that we need charged shockwaves to bound $\alpha_i$. If instead we take $m_0 \to 0$, the bound becomes 
\begin{align}
     \alpha_i  \ > \ \frac{\rho^2 }{48 M_\text{P}^2} \, .
\end{align}
The strongest possible bound comes from the largest possible value of $\rho$, which is $\rho \sim 1 / m_\text{IR}$. Recall also the expectation from the dispersion relation arguments that $\alpha_i \sim 1 / M^4$. Putting this together, we find a parametric bound
\begin{align}
 M^4 \lesssim  M_\text{P}^2  m_\text{IR}^2 \, .
\end{align}
Surprisingly, this is the same as the Cohen-Kaplan-Nelson bound \cite{Cohen:1998zx}, which uses the fact that in a theory with gravity, the entropy in a region scales with the area to derive a relation between the UV and IR  cutoff scales. Our argument is completely independent, and in particular does not assume anything about black hole physics.

\section{More shockwaves}
\label{sec:furthershockwaves}

Here we will list a number of  shockwave solutions which exist in the literature. We will mainly discuss gravitational shockwaves, of which there are a surprisingly large number. There are no new results in this section; we only present the known shockwave solutions for future reference and general interest.

\subsection{Higher-dimension RN}

First let us derive the higher-dimensional charged shockwaves in the same way that we did for the higher-dimensional neutral ones. Charged shockwaves in general dimensions were first given in \cite{Lousto:1988ua}, and have been further considered in \cite{Ortaggio:2006gh, Yoshino:2006dp, Arefeva:2009pxq}. We will derive them ourselves to ensure consistency with our choice of normalizations, which are consistent with the Lagrangian given in~(\ref{eq:EMlagrangian}). 
We start with Reissner-Nordstr{\"o}m in general dimensions, 
\begin{align}
    & ds^2 = g(r) dt^2 + g(r)^{-1} dr^2 + r^2 d\Omega^2_{d-2} \, , \\
    & \qquad g(r) = 1 - \frac{2m}{r^{d-3}} + \frac{q^2}{r^{2d-6}} 
    \, , \\
    & \qquad A = \sqrt{\frac{2(d-2)}{d-3}} \frac{q}{r} dt \, ,
\end{align}
where the parameters $m$ and $q$ are related to the physical mass and charge by
\begin{align}
    M = \frac{\Omega_{d-2}}{16 \pi} 2(d-2) \, m, \qquad Q = \frac{\Omega_{d-2}}{16 \pi} \sqrt{2 (d-2)(d-3)} \, q \, , 
\end{align}
with the sphere volume factor $\Omega_{d-2} = 2 \pi^{\frac{d-1}{2}} / \Gamma(\frac{d-1}{2})$. After a suitable change of coordinates the metric takes the isotropic form,
\begin{align}
    ds^2 = -\left(1+ \frac{m^2-q^2}{4 r^{2d-6}}\right)^2\frac{dt^2}{R^2} + R^{\frac{2}{d-3}} (dr^2 + r^2 d\Omega_{d-2}^2) \, , 
\end{align}
and the gauge field becomes
\begin{align}
    A = \sqrt{\frac{2(d-2)}{d-3}} \frac{q}{R} dt \, , 
\end{align}
with the new function 
\begin{align}
    R = \left(1 +  \frac{m}{r^{d-3}}+ \frac{m^2-q^2}{4 r^{2d-6}}\right) \, .
\end{align}
After applying a boost, we find that the metric takes the form
\begin{align}
    ds^2 = \eta_{\mu \nu} dx^\mu dx^\nu + h(u, x_i) du^2 \, ,
\end{align}
with 
\begin{align}
    h(u, x_i) = \delta(u) \sqrt{\pi} \left( m_0   \frac{ (d-2) \Gamma\left( \frac{d-4}{2}\right)}{ \rho^{d-4} \Gamma\left( \frac{d-1}{2}\right)}  - q_0^2  \frac{ (2d-5)  \Gamma\left( \frac{2d-7}{2}\right)}{ 2 \rho^{2d-7} \Gamma\left( d-2\right)} \right) \, .
\end{align}
The time delay of a scalar can be read off from the boosted background. For example, we can compute the time delay of a charged scalar in 5 dimensions by plugging in $d = 5$. The result is 
\begin{align}
    \Delta v =   m_0 \frac{3 \pi}{\rho} - q_0^2\frac{5 \pi}{8 \rho^3}  \, .
\end{align}
We have verified this result, and in particular, that the scalar's mass and charge decouple, by computing the time delay directly using the equations of motion. 

\subsection{Spinning shockwaves}

The case of Kerr black holes is significantly more complicated. The naive strategy of boosting the metric was first pursued in \cite{Lousto:1989ha}, which considered boosts along the direction of angular momentum, and computed the infinite boost limit, scaling $m \to m_0 / \gamma$ and keeping constant the Kerr parameter $a$ defined by the angular momentum via $J = a M$. The authors computed the limit by expanding the metric in a series in $(a^2 / \rho^2)$. However, subsequent work by Balasin and Nachbagauer pointed out issues with this procedure. The main problem, described in \cite{Balasin:1994tb}, is an ambiguity in the infinite-boost limit, which introduces divergences whose regularization is ambiguous in the absence of further physical input. The strategy pursued in \cite{Balasin:1994tb, Balasin:1995tj, Barrabes:2003up} is to consider instead the infinite boost limit of a delta-function stress tensor which supports the Kerr geometry \cite{Balasin:1993kf}, and then solve the Einstein equation for the boosted stress tensor. A later, simpler analysis \cite{Barrabes:2003up} confirmed the result of Balasin and Nachbagauer by computing the infinite boost limit of the Riemann tensor. The result of these papers is the following: for boosts in the direction of the angular momentum, the profile function $f$, defined by
\begin{align}
    ds^2 = \eta_{\mu \nu} dx^\mu dx^\nu  + f(\rho) \delta(u) \, du^2 \, , 
\end{align}
becomes
\begin{align}
    f(\rho) = - 4 G m_0 \log (\rho^2) +  4 G m_0 \theta(a - \rho) \left( 2 \log \frac{\rho}{a + \sqrt{a^2 - \rho^2}} +  \sqrt{1 - \frac{\rho^2}{a^2}} \right) \, .
    \label{eq:BNparallel}
\end{align}
For $\rho>a$, this is exactly the same as the Aichelberg-Sexl solution \cite{Aichelburg}, so no spin effects appear\footnote{Later work \cite{Yoshino:2004ft} showed that for the higher-dimensional black holes with one angular momentum, there is no ambiguity associated with boosting the metric directly. Computing this metric for general $d$ and then setting $d = 4$ at the end yields the result (\ref{eq:BNparallel}) of Balasin and Nachbagauer.}. For a boost along $x$, orthogonal to the direction of rotation $z$, we have
\begin{align}
   f(\rho) =  - 4 G m_0 \log \left( (y+a)^2 + z^2 \right) \, .
\end{align}
In fact, this metric also contains no spin effects because it can be changed back to the AS solution with the simple coordinate transformation $y \to y - a$. 

Another solution presented in the literature is the one of Ferrari-Pendenza \cite{Ferrari:1990tzs}, computed by boosting the Kerr metric,
\begin{align}
    f(\rho) = - 4 G m_0 \log \left( \rho^2+a^2  \right)\, ,
\end{align}
which does appear to have non-trivial spin effects. 
This case was recently addressed in \cite{Adamo:2021rfq} using the relation \cite{tHooft:1987vrq} between the classical solution of a particle traveling through a shockwave and the eikonal phase of the amplitude for a particle interacting with the source particle. The authors confirm the results of \cite{Balasin:1994tb, Balasin:1995tj}, and explain that the Ferreari-Pendenza metric is actually associated with a stress tensor that is completely delocalized in the $y-z$ plane, rather than a delta-function source as claimed in \cite{Ferrari:1990tzs}. For other recent studies involving boosted Kerr solutions, see \cite{Duenas-Vidal:2022mxb, Frolov:2022rbv}.

\subsection{Shockwaves on black holes}

Among the key early works on shockwaves are those of Dray and 't Hooft, 
\cite{Dray:1984ha}. 
It was known that shockwave geometries in 4d flat space are equivalent to two half-spaces glued together along the $u = 0$ plane front. Dray and 't Hooft derived the general conditions under which this construction, \textit{i.e.} a diffeomorphism, can be used to add a shockwave to a given background. For a solution of the vacuum field equations of the form
\begin{align}
    ds^2 = 2 A(u,v) du dv + g(u,v) h_{ij}(x_i) dx^i dx^j \, ,
\end{align}
a shockwave can be added by a coordinate shift $v \to v + f$ if, at $u = 0$,
\begin{align}
    \partial_v A = \partial_v g &= 0 \, , \\
    \frac{A}{g} \Delta^2 f + \frac{\partial_{uv} g}{g} f &= 32 \pi p A^2 \delta(\rho) \, ,
\end{align}
where the laplacian $\Delta^2$ is defined relative to the 2d-metric $h_{ij}$. The first condition is satisfied by Minkowski, where $A = -1/2$ and $g = 1$, leading to the equation
\begin{align}
    \Delta^2 f = - 16 \pi p \,\delta(\rho) \, ,
\end{align}
determining that $f \sim \log \rho$. Furthermore, the Schwarzschild black hole written in Kruskal coordinates is
\begin{align}
    ds^2 = -\frac{32m^3}{r} e^{-r/2m} du dv + r^2 (d\theta^2 + \sin^2 \theta d\phi^2)\, ,
\end{align}
with $ r$ defined by
\begin{align}
    uv = \left( \frac{r}{2m} - 1 \right) e^{-r/2m} \, .
\end{align}
This metric also satisfies the Dray 't Hooft conditions, so the shockwave metric on the black hole background has the profile function
\begin{align}
    f = \kappa \int_0^\infty ds \frac{ \cos \tfrac{1}{2} \sqrt{3} s}{\sqrt{2(\cosh s - \cos \theta)}} \, ,
\end{align}
where $\kappa = 512 m^4 p e^{-1}$, and $e$ is simply Euler's number. This result applies to a shockwave arranged at $\theta = 0$; subsequent work \cite{Dray:1985yt} extended the result to a spherical shell falling into the black hole. These ideas culminated in 't Hoofts black hole S-matrix \cite{tHooft:1996rdg}, an important precursor to the connection between black hole thermodynamics and chaos. See, for instance, section 2 of \cite{Polchinski:2015cea}.

\subsection{Shockwaves in curved backgrounds} There is a large literature on shockwave solutions in spaces with a non-zero cosmological constant, which we will only mention briefly. See also \cite{Galante:2023uyf} for a nice review focusing mainly on dS. The AS solutions were first generalized to (A)dS by Hotta and Tanaka \cite{Hotta:1992qy, Hotta2} in four dimensions. Later these results were generalized by Sfetsos \cite{Sfetsos:1994xa} to include shockwaves added to a number of more complex backgrounds including those with matter or event horizons, \textit{\`a la} Dray and 't Hooft. Horowitz and Itzhaki \cite{Horowitz:1999gf}  considered higher dimensions and also showed how the CFT duals of ultraboosted black hole states can be used to resolve some apparent paradoxes about bulk and boundary causality. 

A particularly simple form for the dS shockwave metric is given by
\begin{align}
    ds^2 = \frac{\ell^2}{(\ell^2 - uv)^2} \left( - 4 \ell^2 du dv + (\ell^2 + uv)^2 d \Omega_{d-2}^2 \right)  - \frac{m_0 \, \ell^2}{2} \delta(u) du^2 \, ,
\end{align}
where $\ell$ is the dS length scale. In contrast to the solutions given above by boosting black holes, this shockwave is completely isotropic in the transverse plane, and thus it does not correspond to AS in the $\ell \to \infty$ limit -- see \cite{Hotta:1992qy} for such a solution. The Einstein tensor plus the cosmological constant times the metric gives the LHS of the equations of motion for this background:
\begin{align}
    \left( E_{\mu \nu} + \frac{(d-1)(d-2)}{2 \ell^2} g_{\mu \nu} \right) dx^\mu dx^\nu \ = \ m_0 \, \delta(u) \, du^2 \, .
\end{align}
Thus, we see that this is the solution sourced by a delta-function $T_{uu} = m_0 \,\delta(u)$. Moreover, the null energy condition requires that $m_0>0$. The analogous solution for AdS can be found by sending $\ell^2 \to -\ell^2$,  
\begin{align}
    ds^2 = \frac{\ell^2}{(\ell^2 + uv)^2} \left( - 4 \ell^2 du dv - (\ell^2 - uv)^2 d \Omega_{d-2}^2 \right)  + \frac{m_0 \, \ell^2}{2} \delta(u) du^2 \, .
\end{align}
This solution likewise is sourced by a stress tensor with energy density $m_0$. 

The striking result of these two solutions is that the null energy condition requires that light passing through an AdS shockwave experiences a time delay, while light passing through the dS shockwave is subject to a time advance. In 3d AdS, such shockwave solutions play a central role in the discussions on chaos in holography \cite{Shenker:2013pqa}. They are also the starting point for the Gao-Jafferis-Wall wormhole \cite{Gao:2016bin}. In that work it is shown that turning on a double trace deformation that couples the two boundary CFTs allows one to get a negative $m_0$ for the 3d AdS shockwaves, yielding a time advance rather than a delay, and rendering the wormhole in the bulk traversable. 

A full account of causality in AdS and flat space is given by the Gao-Wald theorem \cite{Gao:2000ga}, which lacks a straightforward interpretation in de Sitter space. One consequence is that the causal diagram in de Sitter grows taller under a generic perturbation, bringing the north and south poles into causal contact (see \cite{Leblond:2002ns} for instance). The effects of dS shocks, and the relevance to chaos and information retrieval, has been considered recently in \cite{Aalsma:2020aib, Aalsma:2021kle, Aalsma:2022swk}. Other recent work \cite{Bittermann:2022hhy} has argued that causality is related to the criterion that the null geodesics at the largest impact parameters propagate ``fastest'', in the sense that they hit the future spacelike boundary with the greatest positive spatial shift relative to the unperturbed background. 

Finally, for completeness we mention the result of \cite{Shenker:2013yza}, who computed the background geometry associated to an infinite number of shocks on a BTZ background in the limit where $n m_0$ (the number of shocks times the momentum of each shock) is held fixed. They found
\begin{align}
    ds = -\ell^2 d\tau^2 + \ell^2 \cos^2 \tau dx^2 + R^2 \left(1 - \sin \tau \log \frac{1 + \sin \tau}{\cos \tau} \right)^2 d\phi^2 \, ,
\end{align}
where $R = 8 \pi G M \ell^2$. The stress tensor corresponding to this geometry is smooth and arises from taking pressureless dust and boosting half in the $x$-direction, and half in the $-x$-direction so that the total momentum is zero. As far as we know, the corresponding geometries for flat space and de Sitter space have not been determined, but perhaps it could be worked out along similar lines.

\subsection{More exotic shockwaves}

For the sake of completeness and fun, let us briefly mention a few other solutions present in the literature. 

\paragraph{Extended objects} As part of their series on ultraboosted solutions, Lousto and Sanchez considered a number of extended objects \cite{Lousto:1990wn}. We will cite some of their results. These are all 4 dimensional, with metrics satisfying the ansatz
\begin{align}
    ds^2 = \eta_{\mu \nu} dx^\mu + dx^\nu + f(y,z) \delta(u) du^2 \, .
\end{align}
The extended solutions include one-dimensional cosmic strings \cite{Vilenkin:1984ib} stretched in the $z$ direction, described by
\begin{align}
    \text{cosmic strings} \qquad  f = 2 \mu_0 |y| \, .
\end{align}
Here the strings are extended in the $z$ direction and the mass density $\mu$ scales with the boost as $\mu = \mu_0 \gamma^{-1}$. 

Likewise, there exist two-dimensional domain walls \cite{Vilenkin:1983ykn}, extended in $y$ and $z$ and boosted in $x$ , 
\begin{align}
    \text{domain wall} \qquad  f = \frac{1}{2} \sigma_0 \rho^2  \, ,
\end{align}
where the mass density scales\footnote{It seems a little strange that the profile function grows like $\rho^2$ when physically this setup should be independent of $\rho$. Nonetheless, we have checked that for $f \sim \rho^n$, the unique solution where the stress tensor is $\rho$-independent comes from $n = 2$.} like $\sigma = \sigma_0 \gamma^{-1}$.

Other interesting  solutions are the so-called ``global monopoles'' \cite{Barriola:1989hx, Harari:1990cz}, which are spherical field configurations associated with the spontaneous breaking of a global symmetry. Such configurations are surrounded by goldstone fields with the energy density decreasing like $r^{-2}$, rather than the usual $r^{-4}$. As a result, the total energy inside a region of size $R$ scales linearly with $R$. A monopole-antimonopole pair attracts each other gravitationally, so such objects are essentially confined. Nonetheless, we can consider their boosted gravitational fields: 
\begin{align}
    \text{global monopole} \qquad  f = \frac{1}{2} e_0 \rho \, ,
\end{align}
where the energy density goes like $ \sim e / r^2$ and the charge has been scaled, $e = e_0 \gamma^{-1}$. 

The common pattern and rule of thumb for these solutions is this: if the energy contained inside a region of size $R$ scales like $R^n$, then the profile function $f$ will go like
\begin{align}
    f \sim x_\perp^n \, .
\end{align}

\paragraph{Gyratons} Another possibility for the ultraboosted limit of spinning objects is to scale the mass so that the energy is finite, but to keep the angular momentum finite. Such objects were introduced by Bonner, who first considered the metric due to a beam of light moving in the $+x$-direction \cite{Bonnor:1969mfs} and then generalized his result to a spinning lightbeam of finite size \cite{Bonnor:1970sb}. They were rediscovered by Frolov and collaborators \cite{Frolov:2005in, Frolov:2005zq}, where the term ``gyratons'' was introduced. See \cite{Podolsky:2014lpa} for an overview, as well as \cite{Frolov:2022rbv} for some recent work.

\paragraph{Boosted Melvin universe} The Melvin universe describes a black hole embedded in a uniform electric field. The ultraboost limit of such solutions was computed in \cite{Ortaggio:2003ri}.

\paragraph{Boosted Taub-NUT} Taub-NUT spacetimes are solutions to the vacuum Einstein equations in 4d with a non-trivial topology. They contain a topological ``NUT'' charge $N$ whose relation to the normal ADM mass is analogous to electric-magnetic duality (see \cite{Bunster:2006rt}). Boosted Taub-NUT solutions with zero mass were computed in \cite{Argurio:2008nb}, and take a particularly simple form,
\begin{align}
    ds^2 = \eta_{\mu \nu} dx^\mu dx^\nu - 8 n_0 \arctan \frac{y}{z} \, \delta(u) du^2 \, ,
\end{align}
where $n_0$ is defined by the scaled NUT charge via $N = n_0 \gamma^{-1}$. The profile function does not depend on the radius $\rho = \sqrt{y^2 + z^2}$, with the $\arctan$ varying smoothly from $- \pi / 2$ to $\pi / 2$ as the angle changes. 

\paragraph{Double copy} Certain boosted solutions in gravity are related to boosted gauge theory configurations via the double copy, see \cite{Saotome:2012vy, Monteiro:2014cda, Bahjat-Abbas:2020cyb}.

\section{Discussion}
\label{sec:Conclusions}

In this paper we have examined how causality of probe particles traveling on shockwave backgrounds can be used to bound EFT operators, in pure field theory and in theories with gravity. We have focused on cases in which the probes are photons, which has allowed us to see how the higher-derivative operators of (Einstein-)Maxwell theory contribute to the time delay. One of the main results is that, due to the rescaling of the charge $q^2 = q_0^2 \gamma^{-1}$, the gauge fields themselves vanish in the infinite boost limit, but various contracted quantities, like the stress tensor due to the gauge field, do not. This fact is crucial to ensure that higher-derivative operators like $(F^2)^2$ can contribute to the time delays. 
For the gravitational case, this results in a time delay in Einstein-Maxwell theory equal to
\begin{align}
   \Delta v= -\frac{3 \pi q_0^2}{M_\text{P}^2 \rho} - 16 \frac{m_0}{M_\text{P}^2} \log\rho   
         +    \alpha_i  \frac{48 \pi q_0^2}{\rho^3} \pm  \,\alpha_3 
 \left(\frac{18 \pi q_0^2}{M_\text{P}^2 \rho^3} - \frac{64 m_0}{M_\text{P}^2 \rho^2} \right),
\end{align}
where $\alpha_i$ refers to $\alpha_1$ or $\alpha_2$, depending on which of the possible polarizations is chosen.
The time delay 
  receives two distinct contributions, one purely from the geometry and one from the higher derivative interactions. Because of its negative sign, the geometry contribution weakens any positivity bounds on the coefficients of the higher derivative terms.

Imposing ``positivity'' of this time delay is slightly ambiguous. Due to the log, the delay is not defined without a reference. This can be chosen to be an IR cutoff scale, allowing for comparison between the causality and dispersion relation bounds of \cite{Henriksson:2022oeu}. 
Alternatively, one can remove the need for a cutoff and get a physical quantity by considering the derivative of the delay with respect to $\rho$. It is not clear that this needs to be positive, but naively imposing that it does implies an interesting parametric relationship between the Planck scale, EFT cutoff scale, and IR cutoff scale which is exactly the same as the Cohen-Kaplan-Nelson bound \cite{Cohen:1998zx}.

\paragraph{Future directions}

Related to the discussion directly above, the direction that is most pressing for the present line of research is to understand exactly what condition causality imposes in four dimensions with gravity. The presence of logarithms and need for an IR cutoff, presumed to be related to the IR divergences present even at tree-level in massless theories,
makes a direct interpretation of the dispersion relation bounds challenging. Our calculation is purely classical so there is hope for a more straightforward argument. We hope to return to this issue in the future.

One very important possible extension of our work is to go beyond the leading EFT operators. It was shown in \cite{CarrilloGonzalez:2022fwg, CarrilloGonzalez:2023cbf} that considering more complicated backgrounds leads to bounds on operators with six and more operators. In this work we have mainly focused on shockwave \textit{solutions}, that is, on backgrounds which themselves solve the equations of motion with particular specified sources, rather than arbitrary field configurations. It would be interesting to relax this constraint and see if more complicated shockwave profiles allow for a larger and more stringent set of bounds.

Extending these analyses to AdS would allow us to make contact with a large body of literature. In fact, it was thinking about how the ANEC bounds higher-derivative operators that initially got us interested in shockwaves. Bulk causality has long been used to bound higher-derivative interactions \cite{Brigante:2007nu, Brigante:2008gz}, though much better bounds are now available -- see \cite{Caron-Huot:2022ugt}. Bulk causality has been understood to be dual to the boundary ANEC, implicitly in \cite{Hofman:2008ar, Hofman:2009ug}, and more explicitly in \cite{Kelly:2014mra}, where a proof of the boundary ANEC from bulk causality was given. For classical gravity, various results about bulk causality require bulk energy conditions, \textit{e.g.} \cite{Gao:2000ga}. One interesting recent discussion \cite{Hartman:2022njz} showed how the focusing theorem in the bulk implies constraints on boundary correlators. It would be interesting to understand if the bulk ANEC or focusing theorem imply constraints on higher-derivative interactions, and if these constraints go beyond what is implied by bulk causality/boundary ANEC.\footnote{However, we have a basic confusion about how to interpret results about the bulk energy conditions in a quantum theory because field redefinitions can mix the photon and graviton, and change which fields are considered to be part of the matter stress tensor. Some papers on higher-derivative corrections to bulk gravitational theories, \textit{e.g.} \cite{Cheung:2018cwt, Cremonini:2019wdk, McPeak:2021tvu} find that physical quantities like the mass and entropy always come in field-redefinition invariant combinations. It would be nice to understand the extent to which field-redefinition invariance, which is very natural when the graviton and photon are on equal footing in the path integral, is relevant to bulk gravitational physics.} Finally, it appears that the commutativity of shockwaves, \textit{i.e.} the requirement that for almost coincident shocks, observables do not depend on which shock appears first, was shown in \cite{Kologlu:2019bco} to imply bounds on the EFT and to be dual to the commutativity of boundary value ANEC operators. It would be very interesting to see if this simple requirement can be used to bound EFT coefficients in a setup like ours.

Another question which puzzles us about the relation between shockwaves and dispersion relations is the following: why do the operators which contribute to the time delay appear to be the same ones which contribute to doubly-subtracted dispersion relations (2SDR)? For example, we have seen with the complex scalar that the charge of the scalar decouples, even on the charged shockwave background. The scalar charge contributes to the scattering amplitude with a single power of $s$ and thus does not contribute to 2SDR -- see \cite{McPeak:2023wmq} for bounds involving the scalar charge using 1SDR. Perhaps we could make the scalar charge accessible through the shockwave analysis by choosing different scaling of the charge, but scaling the charges down more gently would likely render the energy of the solution infinite. 
So perhaps the shockwave analysis is a hint that in theories with long range forces, pure 1SDRs cannot be valid. Relatedly, it would be good to repeat our analysis for charged scalars by adding higher-derivative couplings. In that case, there is a non-trivial two-derivative coupling, and it would be interesting to see in what scenarios it contributes to the time delay. A sharp question is: what scalings are needed to make the two-derivative coefficients, and the scalar charge, contribute to the time delay, and is there any physical scenario where these scalings are justified?

Another question which would be interesting to address is to examine time delays caused by objects with non-zero rest mass and rest charge. Computing the Shapiro delays for light on various backgrounds is, of course, a common research problem. These cannot be computed with the technology used in this paper because the infinite boost limit provides drastic simplifications that make the computation of the time delay much more tractable. It is hard to imagine computing the time delays for the higher-derivative equations of motion on a full black hole background (though see \cite{Papallo:2015rna} for a computation along these lines). The infinite boost limit requires that we scale down the total mass and charge to keep the energy finite, which prevents us from computing, for instance, the time delay of light moving past an electron. Going beyond the infinite boost limit would allow for many more physically interesting results to be computed. For instance, what is the relation between the extremality ratio $q / m$ of the source, and the success/failure of the resulting causality bounds to imply the black hole weak gravity conjecture \cite{Arkani-Hamed:2006emk, Kats:2006xp}?
We leave these questions to future work.

\section*{Acknowledgments}

We would like to thank Simon Caron-Huot, Clifford Cheung, Tom Hartman, Johan Henriksson, Julio Parra-Martinez, Andrew Tolley and Alessandro Vichi for a number of useful conversations and correspondences. This work has received funding from the European Research Council (ERC) under the European Union's Horizon 2020 research and innovation program (grant agreement no.~758903). 
The work of S.C. and Y.T. is supported in part by the NSF grant
PHY-2210271. This work benefited greatly from participation in the program “Bootstrapping Quantum Gravity” at the Kavli Institute for Theoretical Physics at UC Santa
Barbara.  S.C. thanks KITP for their hospitality. 
This research was supported in part by grant NSF PHY-1748958 to the Kavli Institute for Theoretical Physics (KITP).

\appendix
\addtocontents{toc}{\protect\setcounter{tocdepth}{1}}

\section*{}
\bibliography{cite.bib}
\bibliographystyle{JHEP.bst}
\end{document}